\documentclass[english,12pt]{article}

\usepackage{amssymb,amsmath,epsfig}

\usepackage[square,comma,sort&compress,numbers]{natbib}
\usepackage{savetrees}
\usepackage{float}
\usepackage{graphicx}
\usepackage[font=footnotesize,labelfont=bf]{caption}
\usepackage{sidecap}
\sidecaptionvpos{figure}{c}
\usepackage{subfigure}
\usepackage{url}
\usepackage{comment}
\usepackage{color}
\usepackage[mathscr]{euscript}

\def\Tc{$T_{\mathrm{c}}$ }
\def\Ic{$I_{\mathrm{c}}$ }

\def\mnras{Mon. Not. R. Astron. Soc.}

\def\nat{Nature}

\def\aap{Astron. \& Astrophys.}

\def\pasp{Pub. Astron. Soc. Pac.}

\def\procspie{Proc. SPIE}

\def\prl{Phys. Rev. Lett.}
\def\prb{Phys. Rev. B}

\newcommand{\beq}{\begin{equation}}
\newcommand{\eeq}{\end{equation}}

\usepackage[LGR,T1]{fontenc}
\usepackage[latin9]{inputenc}
\usepackage{subscript}



\usepackage{babel}
\usepackage{setspace}
\title{Demonstration of ultra-low noise equivalent power using a longitudinal proximity effect transition-edge sensor}
\author{Peter C. Nagler, John E. Sadleir and Edward J. Wollack}
\date{NASA/Goddard Space Flight Center, Greenbelt, MD USA}

\begin{document}
\maketitle

\begin{abstract}
\noindent
Future far-infrared astronomy missions will need large arrays of detectors with exceptionally low noise-equivalent power (NEP), with some mission concepts calling for thousands of detectors with NEPs below a few $\times 10^{-20}$ W/$\sqrt{\mathrm{Hz}}$. Though much progress has been made toward meeting this goal, such detector systems do not exist today. In this work, we present a device that offers a compelling path forward: the longitudinal proximity effect (LoPE) transition-edge sensor (TES). With a chemically-stable and mechanically-robust architecture, the LoPE TES we designed, fabricated, and characterized also exhibits unprecedented sensitivity, with a measured electrical NEP of $8 \times 10^{-22}$ W/$\sqrt{\mathrm{Hz}}$. This represents a >100x advancement of the state-of-the-art, pushing TES detectors into the regime where they may be employed the achieve to goals of even the most ambitious large and cold future space instruments.  
\end{abstract}

%\tableofcontents

%\newpage

%\pagenumbering{roman}

\pagenumbering{arabic}

\section{Introduction}

Experiments employing incoherent low temperature detectors have made significant contributions to astronomy from the X-ray through the millimeter-wave. Offering superior noise performance compared to detectors that operate at higher temperatures, the earliest widely-deployed detectors of this type were based on doped semiconductors cooled below their metal-insulator transition, where a temperature-dependent conduction mechanism (variable range hopping) enables the material to act as a thermometer sensitive to deposited energy or power. Space-based instruments flying on {\sl COBE} \cite{Matheretal1993}, {\sl Planck} \cite{Lamarreetal2010}, and {\sl Astro-H/Hitomi} \cite{Kelleyetal2016} used detectors based on this technology. More recently, both equilibrium and non-equilibrium superconducting detectors have largely replaced semiconducting detectors, offering higher sensitivity, simpler fabrication, and multiplexed readout schemes for large numbers of detectors.  Many thousands of superconducting detectors are currently deployed from ground (e.g., SCUBA-2 \cite{Hollandetal2013}, BICEP-3 \cite{BICEP32016}, ACTpol \cite{Thornton2016}, and Subaru \cite{Walteretal2020}), aircraft (e.g., HAWC+\cite{Harper2018}), and balloon (e.g., Spider \cite{Bergmanetal2018}, EBEX \cite{EBEX2018}, PIPER \cite{PIPER2020}, and PICTURE-C \cite{Mendilloetal2020}) platforms. Future space telescopes - particularly those with large and/or cooled apertures - are likely to require the use of superconducting detectors to achieve background-limited sensitivity.

Despite considerable effort, demonstrating detector systems achieving the ultimate sensitivity and scale required for these future missions remains elusive. For example, the Origins Space Telescope concept ({\sl Origins}), which features a 6.5 m primary mirror cooled to <5 K, requires more than 10,000 far-infrared (FIR) detectors with noise-equivalent power below $3 \times 10^{-20}$ W/$\sqrt{\mathrm{Hz}}$ in order to make background-limited observations of far-infrared (FIR) spectral lines \cite{Origins2019}. Superconducting transition-edge sensor (TES) bolometers have been identified by the mission as the baseline detectors capable of meeting this requirement, but published performance is still an order of magnitude away \cite[e.g.,][]{Weietal2008,Beyeretal2010,KarasikandCantor2011,Khosropanahetal2016,Suzukietal2016}. Microwave kinetic inductance detectors (MKIDs) \cite{Dayetal2003} - another superconducting detector technology - also show promise, but the best measured MKID performance is similar to that achieved by TESs \cite{deVisseretal2014}. Recently quantum capacitance detectors (QCDs) have emerged as a technology able to meet the sensitivity requirements of missions like {\sl Origins} \cite{Echternachetal2018}, but uniform arrays of the scale envisioned for future space telescopes ($>10^{4}$ pixels), along with a practical large-format readout scheme, have yet to be shown.

In this work, we present a TES bolometer with a measured electrical NEP of $<1 \times 10^{-21}$ W/$\sqrt{\mathrm{Hz}}$ in the signal band from 1 Hz to 1 kHz, representing a >100x improvement over the state-of-the-art. The device is fabricated from chemically-stable materials on a mechanically-robust solid substrate and uses the longitudinal proximity effect (LoPE) from attached Nb leads to induce superconductivity in a Au sensor film. We refer to this new type of TES as a ``LoPE TES.'' We present both measurements and a complete theoretical model that explain the breakthrough performance of this device, and discuss the implications of its sensitivity for astronomy and other applications.

\section{The longitudinal proximity effect (LoPE) TES}

\begin{figure}
    \centering
    \includegraphics[width=0.49\textwidth]{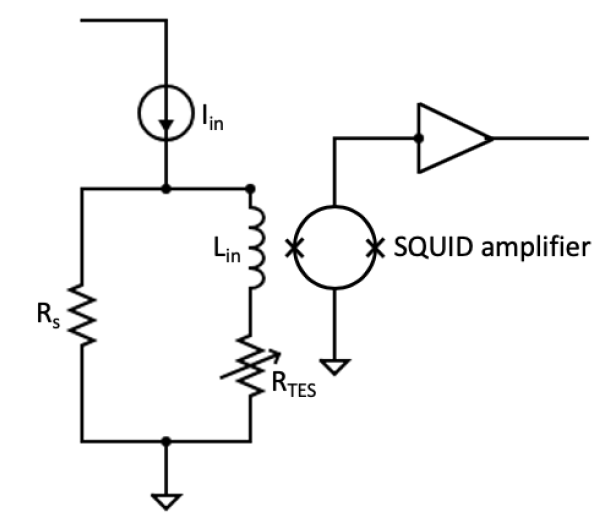}
    \includegraphics[width=0.49\textwidth]{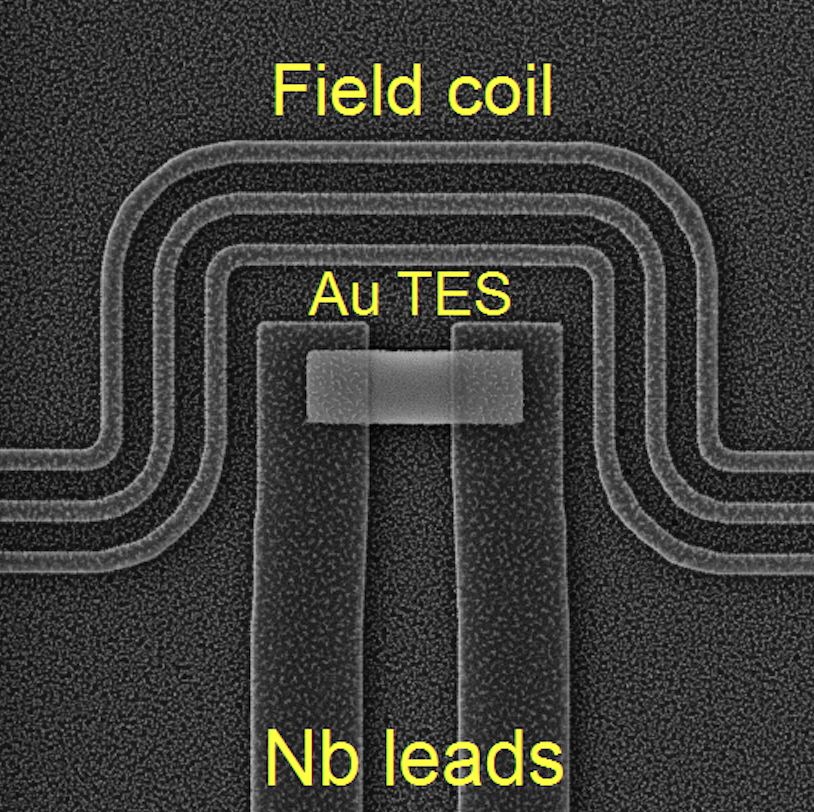}
    \caption{Left: Basic TES bias and readout circuit. Changes in the TES resistance $R_{\mathrm{TES}}$ change the current flowing through the input coil to the SQUID amplifier, which serves as a sensitive ammeter. Right: SEM of a representative detector. The leads are Nb and the TES sensor is Au. This particular sensor is 6 $\mu$m $\times$ 6 $\mu$m. The on-chip Nb field coil is capable of applying a few 10s of mT before exceeding the critical current of the Nb traces. In the measurements presented in this paper, it is used to null the $\sim 10$ $\mu$T ambient magnetic field in the detector package and also to measure the dependence of critical current on applied field. An off-chip coil could also be used for this purpose, but the on-chip version is useful for laboratory characterization.}
    \label{fig:TES_circuit}
\end{figure}

Superconductivity is a thermodynamic phase that exists below a critical temperature $T_{\mathrm{c}}$, current $I_{\mathrm{c}}$, and magnetic field $B_{\mathrm{c}}$.  Transitioning through a material's normal phase (non-superconducting) to the superconducting phase, while carrying a finite DC bias current, the electrical resistance undergoes an abrupt decrease from its normal state resistance $R_N$ to zero. Superconducting transition-edge sensors (TESs) exploit this steep resistive transition to precisely measure deposited energy (as in TES quantum calorimeters) or deposited powers (as in TES bolometers).  The change in TES resistance results in a change in current through the device, measured by a superconducting quantum interference device (SQUID) amplifier employed as an ammeter (Fig. \ref{fig:TES_circuit}).

In Sadleir {\sl et al.} \cite{Sadleiretal2010}, measurements of a bilayer TES's critical current versus temperature, magnetic field, and over a range of geometries showed excellent agreement with a Ginzburg-Landau (GL) model for the TES and the hallmarks of a superconducting Josephson weak-link: exponential temperature dependence, Fraunhofer-like oscillations with applied magnetic field, and a Josephson current-phase relation \cite{Josephson1962,Josephson1964,Josephson1965}.  This GL model and microscopic theoretical treatment of the system both showed that superconducting order from the higher \Tc superconducting leads diffused longitudinally into the variable resistor region of the TES effectively raising the TES's transition temperature.  This longitudinal proximity effect was observed over remarkably long distances, in excess of 1000 times the electron mean free path $\ell_{\mathrm{mfp}}$.  

These results point to the intriguing possibility of a LoPE TES where the sensor need not be a superconductor at the device's operating temperature and can even be made of materials that are not superconducting at any temperature \cite{Sadleir2013b,Sadleir2014b}.  This greatly expands the TES material design space and provides opportunities to extend state-of-the-art performance for a wide range of applications, enabling materials and designs that can offer combinations of reduced noise, higher responsivity, higher speed, higher quantum efficiency, and improved spectral resolving power.  Additionally, it enables making smaller sized TESs. This can both reduce noise and allow higher density arrays across the electromagnetic spectrum. Importantly it can offer a simplified fabrication procedure with fewer layers and steps.

Here we present the first results on a Au LoPE TES.  The device is designed such that quantum wavefunction governing the superconducting pairs in the Nb leads penetrates into the Au region that constitutes the variable resistor. For a given Au film composition and thickness, and for given interface properties between Au and Nb, the \Tc of the LoPE TES depends on the longitudinal separation $\ell$ between the Nb leads. In this case $\ell$ is chosen to give an operating temperature below 150 mK.

\section{Device description and experimental setup}

The detector employed is a TES that uses a Au sensor film connected to Nb leads. The device was fabricated using standard microfabrication techniques and optical photolithography.  This material combination was chosen for its relative ease of fabrication, chemical stability, and ability to achieve high sensitivity on solid substrates. Additionally, the properties of the material constituents are well understood in thin film form. Figure \ref{fig:TES_circuit} shows a scanning electron micrograph (SEM) of a representative detector.

\begin{figure}
    \centering
    \includegraphics[width=0.49\textwidth]{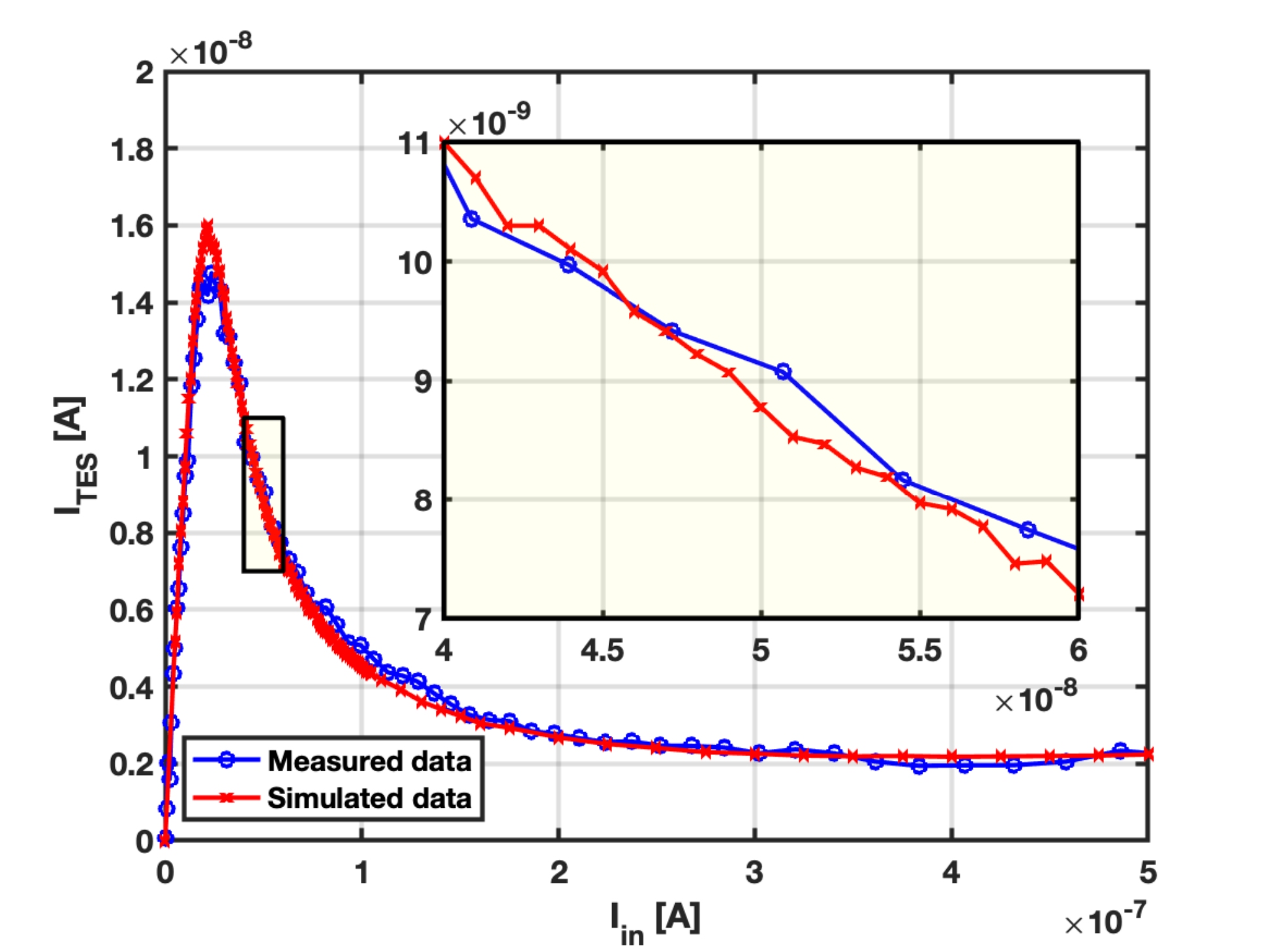}
    \includegraphics[width=0.49\textwidth]{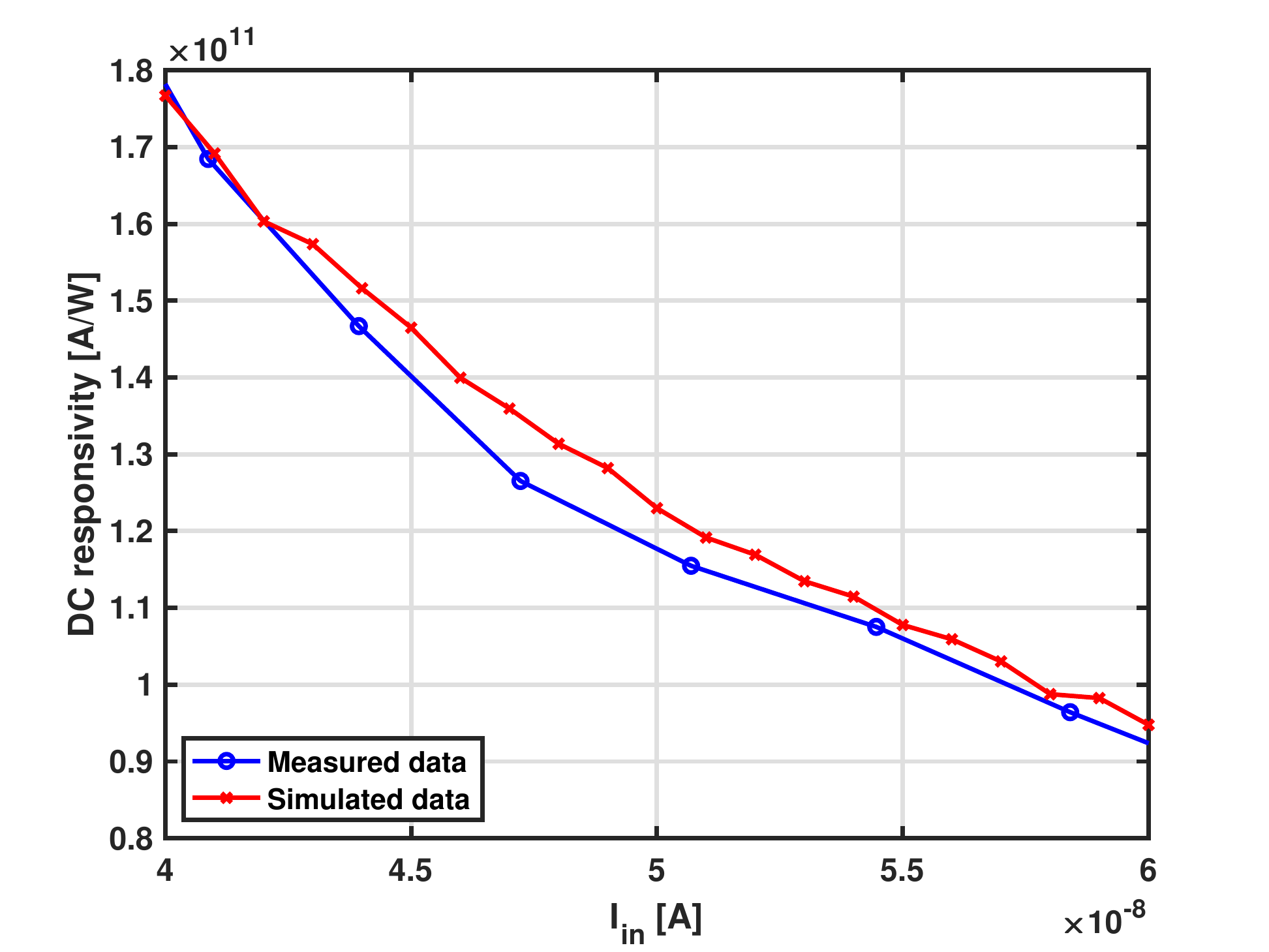}
    \caption{Left: Measured and simulated current-voltage (I-V) curves at 135 mK, showing excellent agreement between measurement and simulation throughout the superconducting-normal transition. The yellow shaded region is expanded in the inset plot. It details the region where we biased the device for noise measurements. Both the measured and simulated data are computed by taking one second averages of the TES current at a given bias point. Right: DC current responsivity of the device computed from the measured and simulated I-V curves at left. This plot details the region where we biased the device for noise measurements. The strong agreement between the measured and simulated I-V curves yields strong agreement in the calculated responsivities.}
    \label{fig:IV}
\end{figure}

The detectors feature on-chip magnetic field coils that use a superconducting Nb winding. The field coils can be employed to null the ambient magnetic field in the test platform or enable electrical measurements taken as a function of applied magnetic field. The devices presented in this paper have measured critical fields (field at which no critical current is measurable) of a few hundred $\mu$T; the coil itself is capable of supplying fields of >10 mT before exceeding the critical current of the Nb windings. Critical current versus field measurements for these devices reproduce the familiar Fraunhofer-like diffraction pattern measured previously in MoAu TESs \cite[e.g., see][]{Sadleiretal2010,Sadleiretal2011} and anticipated in weak-link TESs \cite{Sadleiretal2013}.

The TESs are read out using series arrays of superconducting quantum interference devices (SQUIDs). Each TES sensor has its own readout channel. The devices are cooled with a two-stage adiabatic demagnetization refrigerator (ADR) backed by a pulse tube cooler. With the TES characterization package and associated wiring installed, the ADR can reach temperatures as low as 36 mK.

\section{Device characterization}

We measured current-voltage (I-V) curves in the dark. The on-chip field coil was employed in these measurements to null the ambient magnetic field normal to the sensor films, estimated to be $\sim 10$ $\mu$T. An example I-V curve measured at 135 mK is shown in Fig. \ref{fig:IV}. From the I-V curves, we determined the device's DC current responsivity $\mathscr{R}$ using the dual of Jones' expression \cite{Jones1953}, where $\mathscr{R}=\left(R-Z\right)/\left(2V\left(R_{s}+Z\right)\right)$. Here $Z$ is the dynamic resistance $dV/dI$, $R$ is the resistance $V/I$, $V$ is the voltage across the TES, and $R_{s}$ is the shunt resistance. The maximum responsivity is $\sim 2\times 10^{11}$ A/W and occurs near the peak of the I-V curve, however estimates of the responsivity in that region are limited by system noise.

\begin{figure}
    \centering
    \includegraphics[height=5.9cm]{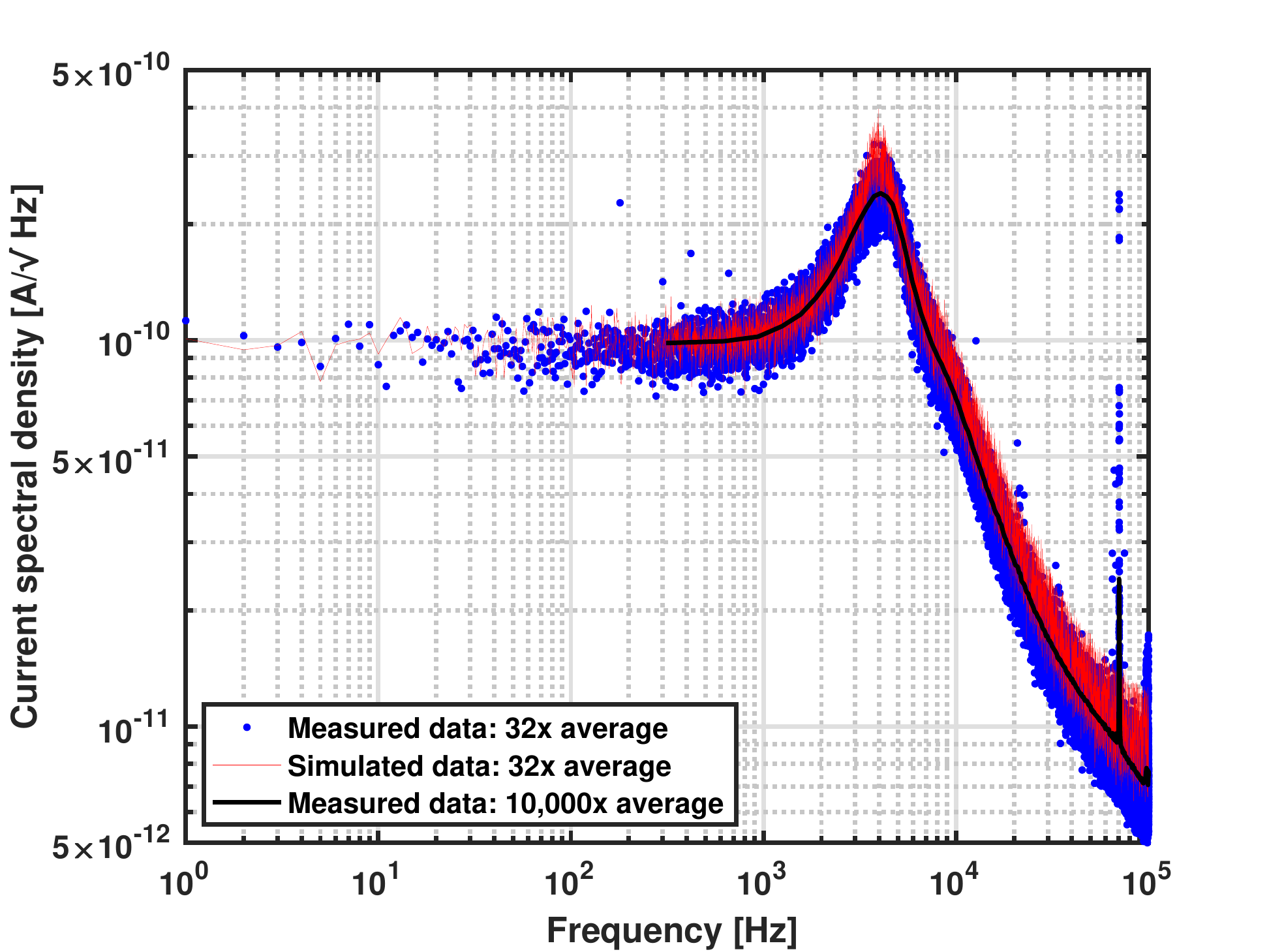}
    \includegraphics[height=5.9cm]{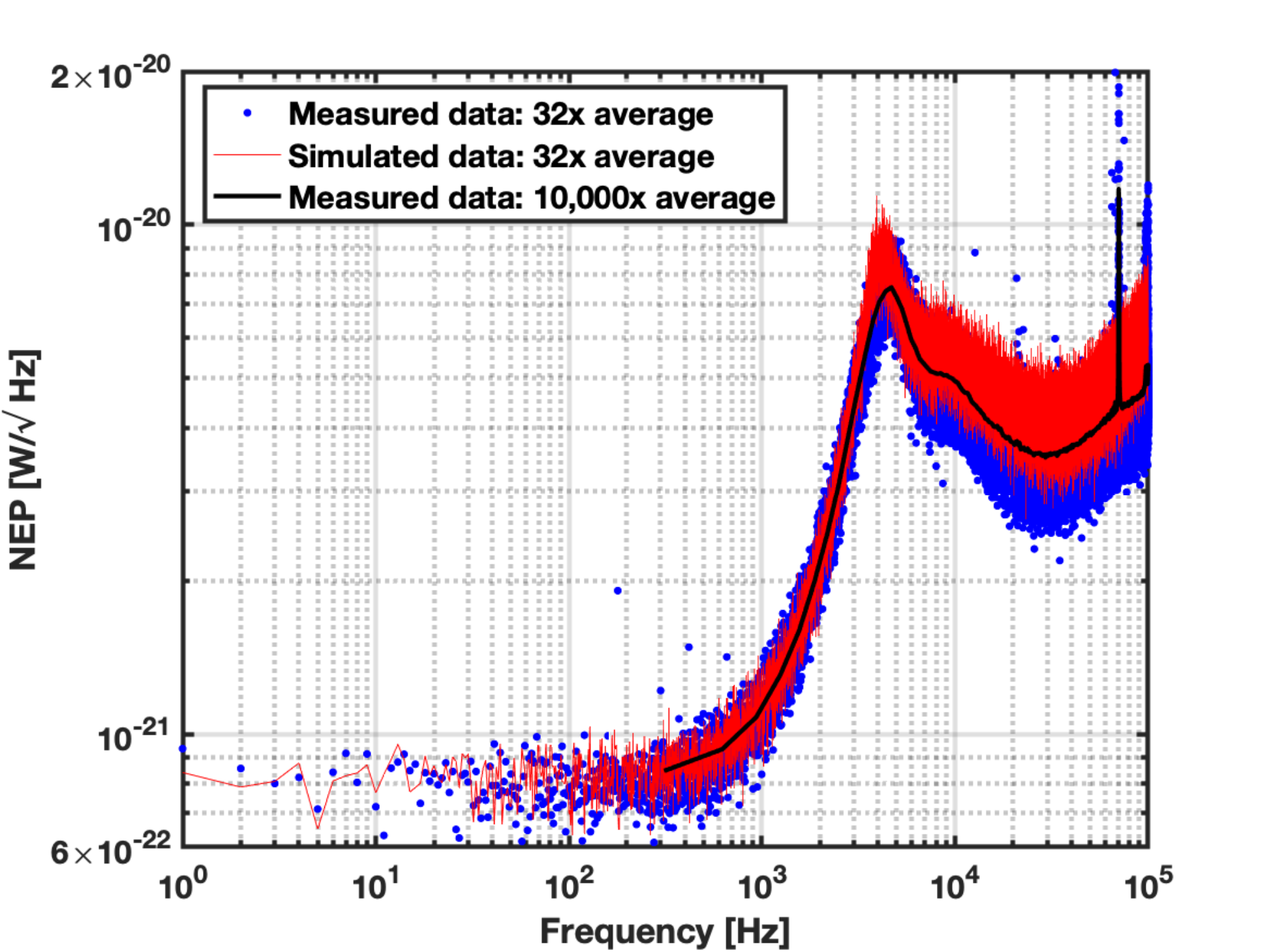}
    \caption{Left: Measured and simulated noise spectra. The measured 32x average current noise trace shown here is the average of 32 consecutive traces, each 1 second in duration. The simulated noise trace is processed in the same way. When the measured traces are instead concatenated, the device's $1/f$ knee is visible at a frequency of $\sim 0.5$ Hz; this is due to the SQUID's contribution. Right: electrical NEP spectra computed using the measured and simulated noise spectra and frequency-dependent electrical responsivity. At the bias point where the data were measured the DC responsivity is $1.2 \times 10^{11}$ A/W. This device achieves a broadband electrical NEP of $8 \times 10^{-22}$ W/$\sqrt{\mathrm{Hz}}$ across the signal band from 1 Hz to 300 Hz, and remains below $1 \times 10^{-21}$ W/$\sqrt{\mathrm{Hz}}$ for frequencies up to 1 kHz.}
    \label{fig:noise_NEP}
\end{figure}

We also measured noise spectra of the devices. To do so, we collected timestreams of the SQUID output at a given bias point. We then took the discrete Fourier transform of a given timestream and computed the power spectral density (PSD). The noise spectrum at the bias point near the maximum measured responsivity is shown in Fig. \ref{fig:noise_NEP}. Shown are the average of 32 PSDs (``32x average'') and 10,000 PSDs (``10,000x average''). Both are processed from the same 32 seconds of timestream data; for $n$ averages, we divide the full timestream equally into $n$ shorter timestreams, calculate the PSD of each, then take their average.

We computed the device's electrical noise-equivalent power (NEP) spectrum from the DC responsivity and noise spectrum. With an approximate single-pole response, the frequency-dependent responsivity $\mathscr{R}\left(f\right)$ is given by $\mathscr{R}\left(f\right) = \mathscr{R}/\left(1+ 2 \pi i f \tau\right)$, where we $\tau$ is the time constant of the device inferred from the noise spectrum and $f$ is the device signal frequency. The NEP spectrum is then obtained by dividing the current spectral density by the modulus of the responsivity, $\left|\mathscr{R}\left(f\right)\right|$. The result measured at 135 mK is shown in Fig. \ref{fig:noise_NEP}.

\section{Theory and discussion}\label{sec:discussion}

We use the resistively shunted junction (RSJ) model \cite{McCumber1968,Stewart1968} to model our device. The RSJ model is commonly used to describe Josephson junctions and Josephson weak-links generally, and has also been previous applied to TESs \cite{SadleirThesis,Kozorezovetal2011,Kozorezovetal2011b,Smithetal2013,Sadleiretal2014}. In this model the TES current is composed of: (1) quasiparticle current through the shunt resistor; and (2) Josephson current $I$ through the weak-link. Equivalently, the variable resistor representing the TES in Fig. \ref{fig:TES_circuit} is replaced by a shunt resistor in parallel with a weak-link. We apply this model to our TES where the weak-link current $I$ is proportional to the sine of the phase, consistent with the GL theory LoPE TES model result \cite{Sadleiretal2011} with $I=I_{\mathrm{c}} \sin{\phi}$, and the voltage is proportional to the time derivative of the phase difference $V=\left(\Phi_{0}/2 \pi\right) \dot{\phi}$, where  $\Phi_0$ is the magnetic flux quantum, and $\phi$ and \Ic are the phase difference and critical current across the weak-link, respectively. We numerically solve the RSJ model in the time domain following the method presented by \citep{waldram1996}. Thermal fluctuation noise at the device temperature of operation is explicitly incorporated in the model. The model outputs TES current as a function of time for a given bias point sampled at 40 MHz. We find excellent agreement between the modeled and the measured data, as seen in both the I-V relation (Fig. \ref{fig:IV}) and the noise spectrum (Fig. \ref{fig:noise_NEP}). To generate these plots, the time domain output of the simulation is analyzed with the same analysis procedure as the measured data. The I-V curve uses the average of a 1 second long timestream, and the noise spectrum is generated by averaging 32 separate noise spectra, with each individual PSD computed from 1 second of time domain data. We point out that the agreement is achieved with no free undetermined parameters, no hidden variables, and with no unknown noise sources. All inputs to the RSJ model are independently measured physical parameters.  

Operating as a bolometer, the useful signal bandwidth of the device is from $\sim 1 - 1000$ Hz. The bandwidth is limited at low frequency by the SQUID's $1/f$ noise and at high frequencies by the resonant feature in the noise spectrum visible near 4 kHz. This feature looks similar to the signature of a slightly underdamped TES biased near an electrothermal instability.  Our bias point is electrothermally stable, overdamped, and not near an instability.  This feature is purely electrical, arising from the Josephson limit cycle frequency. Its shape and fundamental frequency are determined by the electrical circuit, bias condition, and thermal fluctuation noise. 

We may also consider the sensitivity of the device when operated as a calorimeter \cite{Moseleyetal1984}. In this case the important figure of merit is the detector's energy resolution (or equivalently spectral resolution).  Under the assumption that the noise is approximately Gaussian-distributed, the following expression relates the full width half maximum (FWHM) energy resolution $\Delta E_{\mathrm{FWHM}}$ to the NEP:
\begin{equation}
\Delta E_{\mathrm{FWHM}}= 2 \sqrt{2 \log{2}} \left(\int_{0}^{\infty} \frac{4}{\mathrm{NEP}(f)^2} \,df\right)^{-1/2},
\end{equation}
where $f$ is signal frequency \cite{IrwinandHilton2005}. Solving this equation using our measured NEP and bandwidth, we find $\Delta E_{\mathrm{FWHM}} \simeq 0.19$ meV. Using spectral resolving power $R = 4$ as the criteria for noiselessly counting single photons \cite{Nagleretal2020}, the calculated $\Delta E_{\mathrm{FWHM}}$ implies that this device has the intrinsic sensitivity to count single 180 GHz photons. 

In addition to reproducing the measured data, the model we developed gives us the ability to predict the ultimate sensitivity of an optimized LoPE TES based on this design. By changing the bias point and/or reducing the operating temperature, we expect the electrical NEP can improve by more than a factor of 10 while retaining a similar signal bandwidth.

\section{Conclusions}
We designed, fabricated, and characterized a LoPE TES that exhibits exceptionally-low electrical NEP, advancing the sensitivity achieved by TESs by >100x. The device's behavior is well explained by a finite-temperature RSJ model, which reproduces both the measured I-V relation and noise spectrum. Importantly, the device presented here is simple to fabricate, chemically stable, and mechanically robust. The use of a sensor material with no known intrinsic \Tc is a paradigm shift in TES design. 

While the readout circuit used here is not multiplexed, the full range of SQUID multiplexing options developed for TES readout can be used with this device. For the envisioned scale of arrays for a mission like {\sl Origins}, we expect microwave SQUID multiplexing \cite{Irwin2004} to be most appropriate, though other techniques (e.g., time-division multiplexing \cite{Irwin2002}) can be used for laboratory characterization of modestly sized arrays.

To confirm the utility of this device as a radiation detector, its optical NEP needs to be implemented and measured. Antenna coupling \cite{MeesandRichards1991} is likely to be the most suitable radiation absorption technique for this device at FIR frequencies. We plan to add antennas to our next generation of devices, and future laboratory measurements will focus on measuring the optical response. We point out that we can meet the sensitivity requirement of a mission like {\sl Origins} even if the optical NEP is degraded relative to the electrical NEP.

Beyond the application of FIR space missions, the intrinsic sensitivity of this device may enable broad utility in applications needing single photon sensitivity at lower energies than has been demonstrated before, such as quantum photonics, quantum information, and quantum sensing -- or in detection of small energy processes more generally, including particle physics and dark matter searches.

\section{Acknowledgements}
This work was funded by the NASA/Goddard Space Flight Center Internal Research and Development (IRAD) program.

\bibliographystyle{ieeetr}

\end{document}